\documentclass[%
 reprint,
superscriptaddress,
 amsmath,amssymb,
 aps,
prb,
 longbibliography,
 lengthcheck,%
]{revtex4-1}

\usepackage{graphicx}
\usepackage{dcolumn}
\usepackage{bm}
\usepackage{hyperref}
\begin{document}

\preprint{APS/123-QED}

\title{Anisotropic Exchange in ${\bf LiCu_2O_2}$}

\author{Z.~Seidov}
\affiliation{Experimental Physics V, Center for Electronic Correlations and Magnetism, University of Augsburg, D-86135 Augsburg, Germany}
\affiliation{Institute of Physics, Azerbaijan National Academy of Sciences, H. Cavid pr. 33, AZ-1143 Baku, Azerbaijan}

\author {T.~P.~Gavrilova}
\affiliation{Kazan E. K. Zavoisky Physical-Technical Institute RAS, 420029 Kazan, Russia}
\affiliation{ Kazan (Volga region) Federal University, 420008 Kazan, Russia}

\author {R.~M.~Eremina}
\affiliation{Kazan E. K. Zavoisky Physical-Technical Institute RAS, 420029 Kazan, Russia}
\affiliation{ Kazan (Volga region) Federal University, 420008 Kazan, Russia}

\author{L.~E.~Svistov}
\affiliation{P. L. Kapitza Institute for Physical Problems RAS, 119334 Moscow, Russia}

\author{A.~A.~Bush}
\affiliation{Moscow Institute of Radiotechnics, Electronics, and Automation, RU-117464 Moscow, Russia}

\author{A.~Loidl}
\affiliation{Experimental Physics V, Center for Electronic Correlations and Magnetism, University of Augsburg, D-86135 Augsburg, Germany}

\author{H.-A.~Krug~von~Nidda}
\email[]{hans-albrecht.krug@physik.uni-augsburg.de}
\affiliation{Experimental Physics V, Center for Electronic Correlations and Magnetism, University of Augsburg, D-86135 Augsburg, Germany}

\date{\today}

\begin{abstract}
We investigate the magnetic properties of the multiferroic quantum-spin system LiCu$_2$O$_2$ by electron spin resonance (ESR) measurements at $X$- and $Q$-band frequencies in a wide temperature range $(T_{\rm N1} \leq T \leq 300$\,K). The observed anisotropies of the $g$ tensor and the ESR linewidth in untwinned single crystals result from the crystal-electric field and from local exchange geometries acting on the magnetic Cu$^{2+}$ ions in the zigzag-ladder like structure of LiCu$_2$O$_2$. Supported by a microscopic analysis of the exchange paths involved, we show that both the symmetric anisotropic exchange interaction and the antisymmetric Dzyaloshinskii-Moriya interaction provide the dominant spin-spin relaxation channels in this material.

\begin{description}
\item[PACS numbers] 76.30.Fc, 75.30.Et, 75.47.Lx
\item[Keywords] electron spin resonance, anisotropic exchange interactions, low-dimensional magnetism, multiferroics
\end{description}
\end{abstract}


\maketitle


\section{Introduction}

Unconventional magnetic ground states and excitations of frustrated quantum-spin chains represent attractive issues in solid-state physics during the last decades.\cite{Mikeska_2004} They appear under a fine balance and partly compensation of competing dominant exchange interactions and are often caused by much weaker interactions
or fluctuations.\cite{Chubukov_1991, Kolezhuk_2000, Kolezhuk_2005, Dmitriev_2008, Amiri_2015}
Typically, frustration in quasi-one-dimensional (1D) chain magnets is provided by competing interactions, if the nearest-neighbor (NN) exchange is ferromagnetic and the next-nearest neighbor (NNN) exchange is antiferromagnetic. Numerical investigations of frustrated chain magnets within different models
\cite{Hikihara_2008, Sudan_2009, Heidrich_2009}
have predicted a number of exotic magnetic phases, such as planar, spiral, or different multipolar phases. Moreover, theoretical studies show that the magnetic phases are very sensitive to inter-chain interactions and anisotropic interactions in the system.\cite{Furukawa_2008,Zhitomirsky_2010,Nishimoto_2010}
There is a number of magnets which are attractive objects for experimental investigations as realizations of 1D frustrated systems like LiCuVO$_4$, Rb$_2$Cu$_2$Mo$_3$O$_{12}$, NaCu$_2$O$_2$, Li$_2$CuO$_2$, Li$_2$ZrCuO$_4$, and CuCl$_2$,
(see for example, Refs.~\onlinecite{Enderle_2005, Hase_2004, Drechsler_2007, Boehm_1998, Drechsler_2007_2, Banks_2009}). Here we concentrate on LiCu$_2$O$_2$ with its fascinating interplay  of competing  exchange interactions both within the Cu$^{2+}$ chains and the zigzag-ladders formed by neighboring two chains.\cite{Masuda_2005}

LiCu$_2$O$_2$ was first discovered in 1990 during the study of Li/CuO electrochemical cells \cite{Hibble90a} and in turn synthesized on purpose for search of new candidates for high-temperature superconductivity.\cite{Hibble90b} After it was characterized as a low-dimensional quantum antiferromagnet in the late nineties,\cite{Fritschij98,Vorotynov} its exotic ground-state properties received large interest and triggered further experimental efforts, revealing a complex phase diagram,\cite{Roessli2001} suggesting a dimer-liquid state,\cite{Zvyagin2002} coexistence of dimerization and long-range order,\cite{Choi2004} as well as helimagnetism.\cite{Masuda2004,Gippius,Mihaly2006} The interest in LiCu$_2$O$_2$ was even stronger intensified by the discovery of its ferroelectric properties in 2007,\cite{Park2007} i.e. it turned out to be a paramount example of a multiferroic due to the correlation between spin helicity and electric polarization.\cite{Seki2008,Yasui2009,Kobayashi2009} Detailed investigations to resolve the phase diagram of LiCu$_2$O$_2$ have been performed by means of magnetization and dielectric measurements \cite{Bush} as well as neutron scattering,\cite{Masuda_2005} electron spin resonance (ESR), and nuclear magnetic resonance (NMR) studies.\cite{Bush2013} Basically, a susceptibility maximum at a temperature $T_{\rm max}=38$\,K typical for a low-dimensional antiferromagnet and two subsequent phase transitions at $T_{\rm N1}=24.5$\,K and $T_{\rm N2}=23$\,K into the spin-spiral structure, where the latter is accompanied by the occurrence of ferroelectricity, characterize the magnetic and electric properties of LiCu$_2$O$_2$ at low magnetic fields.

In this paper, we report the results of an ESR study of LiCu$_2$O$_2$ single crystals in the paramagnetic regime. This study is performed in order to obtain information on the anisotropic exchange interactions in this material. The knowledge of the anisotropic exchange parameters is important for the interpretation of the magnetic and magneto-electric properties of LiCu$_2$O$_2$ in the magnetically ordered phase.
Previous ESR experiments revealed a single exchange-narrowed Lorentz-shaped absorption line with $g$ values $g_c \approx 2.22$ and $g_a \approx g_b \approx 2.0$ at a microwave frequency of 9\,GHz and $T \gg T_{\rm max}$ as well as $g_c \approx 2.29$ at 227\,GHz.\cite{Vorotynov,Zvyagin2002} The ESR linewidth $\Delta H$ was found to amount more than 1\,kOe at room temperature and to diverge to low temperature on approaching magnetic order with a critical behavior $\Delta H \propto (T-T_{\rm crit})^{(-p)}$ with $T_{\rm crit} = 30$\,K and $p = 1.28$ or 1.35 for $H || c$ or $H \perp c$, respectively, at 9\,GHz and $T_{\rm crit} = 23$\,K and $p = 0.58$ for $H || c$  at 227\,GHz.\cite{Vorotynov,Zvyagin2002} So far the discussion and analysis of the paramagnetic resonance remained on a qualitative level.

Here we present a quantitative analysis of the angular dependence of the paramagnetic resonance linewidth in LiCu$_2$O$_2$ to determine the anisotropic exchange parameters. For this purpose ESR is the method of choice, because the anisotropy of the line broadening is extremely sensitive to anisotropic interactions, while the isotropic exchange contributions only result in a general isotropic narrowing of the ESR signal.
While previous ESR studies have been limited by twinning of the crystals, our present investigations are performed on high-quality untwinned single crystals, which is an essential precondition to determine the anisotropy unequivocally. Based on our ESR data we will show that besides the symmetric anisotropic exchange interaction, the antisymmetric Dzyaloshinskii-Moriya (DM) interaction substantially contributes to the linewidth, and we will suggest a possible DM exchange path.

\section{Crystal structure and exchange interactions}

\begin{figure}[b]
\includegraphics[width=0.9\columnwidth]{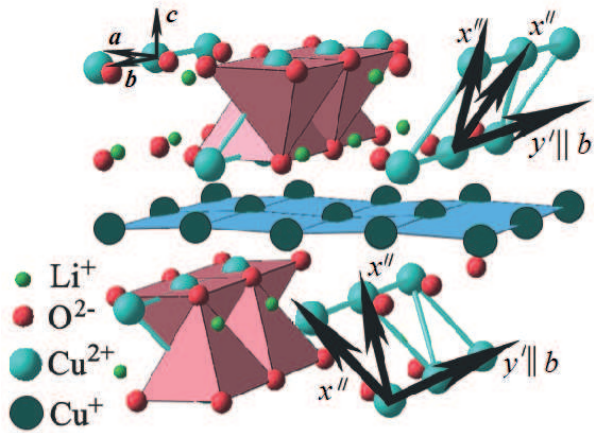}
\caption{(Color online) Orthorhombic crystal structure of LiCu$_2$O$_2$. The spin ladders along the $b$ axis are formed by edge-sharing pink Cu$^{2+}$O$_5$ pyramids. The light blue Cu$^+$ planes separate the structure into layers perpendicular to the $c$ axis. Black arrows indicate the crystallographic coordinates $a,b,c$, the intra-chain local coordinate $y'$, as well as the four possible intra-ladder coordinates $x''$ to describe the local anisotropic exchange tensors.}\label
{LiCu2O2_struct}
\end{figure}

LiCu$_2$O$_2$ crystallizes within an orthorhombic structure (space group $Pnma$). The lattice constants at room
temperature are given by $a=5.726$\,{\AA}, $b=2.8587$\,{\AA}, and $c=12.4137$\,{\AA}.\cite{Berger} Besides four nonmagnetic Li$^{+}$ cations its unit cell contains four monovalent nonmagnetic cations Cu$^+$ (electronic configuration $3d^{10}$) and four divalent cations Cu$^{2+}$ ($3d^9$) with spin $S = 1/2$. Each magnetic Cu$^{2+}$ ion is surrounded by five oxygen ions forming slightly distorted pyramids. Thus, all Cu$^{2+}$ ions are structurally and magnetically equivalent, because the corresponding oxygen pyramids differ from each other by a $180^{\circ}$ rotation, only. There are two linear Cu$^{2+}$ chains in the crystal structure of LiCu$_2$O$_2$, which propagate along the $b$ axis and form a zigzag-ladder like structure as indicated in Fig.~\ref{LiCu2O2_struct}. The ladders are isolated from each other by both Li$^+$ ions in the \emph{ab}~plane and layers of nonmagnetic Cu$^+$ ions along the \emph{c}~direction. The distance between the magnetic nearest-neighbor Cu$^{2+}$ ions along the chains amounts 2.869\,{\AA}, and the spacing between the next-nearest neighbor Cu$^{2+}$ ions (between the chains in one ladder) is about 3.10\,{\AA}.
The unit-cell parameter $a$ is approximately twice the unit-cell parameter $b$. Consequently, LiCu$_2$O$_2$ crystals, as a rule, are characterized by twinning due to formation of crystallographic domains
rotated by 90$^{\circ}$ around their common crystallographic $c$ axis.\cite{Bush}

\begin{figure}
\includegraphics[width=0.9\columnwidth]{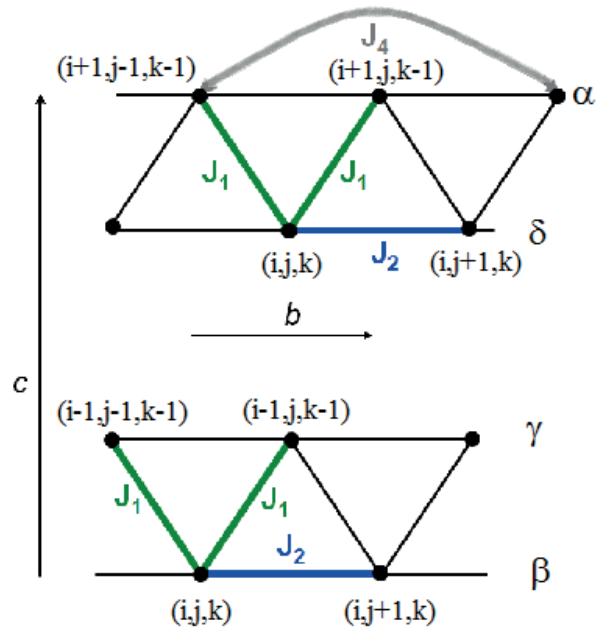}
\caption{(Color online) A schematic view of the exchange interactions between
magnetic Cu$^{2+}$ ions in LiCu$_2$O$_2$. Indices $i$, $j$, and $k$ run over the Cu$^{2+}$ ions along the crystallographic axes \emph{a}, \emph{b}, and \emph{c}, respectively. Symbols $\alpha$, $\beta$, $\gamma$, and $\delta$ denote the four Cu$^{2+}$ chains that form the two ladders $\alpha\delta$ and $\beta\gamma$.}\label
{LiCu2O2_schematic_view}
\end{figure}

The exchange constants of the quasi-one-dimensional helimagnet LiCu$_2$O$_2$ were determined by T.~Masuda \textit{et al.}\cite{Masuda_2005} in a single-crystal inelastic neutron-scattering study. Based on these experiments the authors investigated the validity of three different exchange models and concluded that in LiCu$_2$O$_2$ the frustration mechanism is rather complex and involves a competition between a combination of antiferromagnetic intra-ladder $J_1$ and ferromagnetic intra-chain $J_2$ exchange interactions against an additional antiferromagnetic long-range intra-chain $J_4$ coupling as illustrated in Fig.~\ref{LiCu2O2_schematic_view}. The three corresponding exchange constants turned out to be of comparable strength, i.e. $J_1=3.2$\,meV, $J_2=-5.95$\,meV, and $J_4=3.7$\,meV. Moreover, a sizable antiferromagnetic inter-ladder exchange $J_\perp = 0.9$\,meV was obtained. \cite{Masuda_2005}
Note that the key features of this exchange model, namely, a ferromagnetic $J_2$ bond and a substantial antiferromagnetic $J_4$ coupling constant, are similar to those of theoretical LDA calculations.\cite{Gippius}

As a further corroboration of this model, it can be inferred that recently Y.~Qi and A.~Du \cite{Qi} adopted the suggestion of Masuda \textit{et al.}\cite{Masuda_2005} about a strong antiferromagnetic 'rung' interaction $J_1$ and a weak inter-ladder coupling $J_\perp$, to explain the fascinating magnetoelectric coupling effects observed in LiCu$_2$O$_2$.
Thus, the present analysis of our ESR results will be based on Masuda's exchange model and on our earlier work\cite{Krug von Nidda,Eremina1} in the related compounds LiCuVO$_4$ and CuGeO$_3$.

\section{Theoretical Background}

Electron spin resonance (ESR) measures the resonant microwave-power absorption at a given frequency $\omega$ due to induced magnetic dipolar transitions between the Zeeman levels of magnetic ions split by an external magnetic field $H$. The resonance condition $\hbar \omega = g \mu_{\rm B} H$ yields the $g$ value, which contains information on crystal-electric field and spin-orbit coupling. Here $\hbar$ denotes the Planck constant $h$ divided by $2\pi$, and $\mu_{\rm B}$ the Bohr magneton. The resonance linewidth $\Delta H$ provides microscopic access to the anisotropic interactions acting on the electron spins.

In general, in the case of sufficiently strong exchange interaction the ESR linewidth can be analyzed in terms of the high-temperature approach $(k_{\rm B}T \gg J)$: \cite{Abragam,Anderson1953,Kubo1954}
\begin{equation}
\label{math/1}
\Delta H=\frac{\hbar}{g\mu_{\rm B}}\frac{M_2}{\omega_{\rm ex}}
\end{equation}
where the second moment $M_2$ is defined by:
\begin{equation}
\label{math/2}
M_2=-\frac{1}{\hbar^2} \, \frac{Tr \, [{\cal H}_{\rm int},S_x]^2}{Tr \, S_x^2},
\end{equation}
The second moment $M_{2}$ and the exchange frequency $\omega_{\rm ex}$ can be expressed via microscopic Hamiltonian parameters ${\cal H}_{\rm int}$. The second moment shows an orientation dependence with respect to the external magnetic field, which is characteristic for the anisotropic interaction responsible for the line broadening. The exchange frequency is defined by the dominating exchange interactions shown in Fig.~\ref{LiCu2O2_schematic_view} as
\begin{equation}
\label{math/33}
\omega_{\rm ex}=\sqrt{2J_{1}^2+2J_{2}^2}.
\end{equation}
In LiCu$_2$O$_2$ the second moment is defined by anisotropic interactions of relativistic nature.
Due to the fact that in the case of spin $S=\frac{1}{2}$ the usually dominating single-ion anisotropy is absent, relativistic interactions of neighboring spins, i.e., anisotropic exchange interactions have to be considered.
Note that the magnetic resonance properties of several spin $S=\frac{1}{2}$ chain compounds,\cite{Zakharov} e.g., LiCuVO$_4$ (Ref.~\onlinecite{Krug von Nidda}) and CuGeO$_3$ (Ref.~\onlinecite{Eremina1}) as well as CuTe$_2$O$_5$ (Refs.~\onlinecite{Eremina, Gavrilova}) have been well explained taking into account the anisotropic exchange interactions.

Thus, the results of our present paramagnetic resonance experiments in LiCu$_2$O$_2$ will be discussed in the frame of the following model Hamiltonian:
\begin{eqnarray}
\label{math/4} {\cal H}_{\rm int} &=& J_2^{n} \, (S_{i,j,k}\cdot S_{i,j+1,k})+ S_{i,j,k} \textbf{J}_\textbf{2}^{\textbf{n}} S_{i,j+1,k} \nonumber \\
&+& J_1^{\alpha\delta} \, (S_{i,j,k}\cdot S_{i+1,j,k-1}) + S_{i,j,k}\textbf{J}_\textbf{1}^{\alpha\delta} S_{i+1,j,k-1} \nonumber \\
&+& J_1^{\alpha\delta} \, (S_{i,j,k}\cdot S_{i+1,j-1,k-1}) + S_{i,j,k}\textbf{J}_\textbf{1}^{\alpha\delta}
S_{i+1,j-1,k-1} \nonumber \\
&+& J_1^{\beta\gamma} \, (S_{i,j,k}\cdot S_{i-1,j,k-1}) + S_{i,j,k}\textbf{J}_\textbf{1}^{\beta\gamma}
S_{i-1,j,k-1} \nonumber \\
&+& J_1^{\beta\gamma} \, (S_{i,j,k}\cdot S_{i-1,j-1,k-1}) + S_{i,j,k}\textbf{J}_\textbf{1}^{\beta\gamma}
S_{i-1,j-1,k-1} \nonumber \\
&+& J_\bot \, (S_{i,j,k}\cdot S_{i+1,j,k})+J_4^{n} \, (S_{i,j,k}\cdot S_{i,j+2,k}) \nonumber \\
&+& \textbf{D}_\textbf{2}^{\textbf{n}} \cdot [S_{i,j,k} \times S_{i,j+1,k}] + \mu_B H \cdot\textbf{g}_\textbf{i,j,k} \cdot S_{i,j,k}
\end{eqnarray}
where $n =\alpha, \beta, \gamma, \delta$ denotes the chains corresponding to Fig.~\ref{LiCu2O2_schematic_view}. The summation over all $i,j,k$ is dropped for brevity.
In this model spin Hamiltonian, we included isotropic and symmetric anisotropic exchange interactions between a few types of ions (see Fig.~\ref{LiCu2O2_schematic_view}): ferromagnetic isotropic intra-chain exchange $J_2$ between nearest Cu$^{2+}$ ions in the chains with the tensor of the anisotropic contribution $\textbf{J}_\textbf{2}$, antiferromagnetic isotropic intra-ladder exchange $J_1$ along the rungs with the tensor of anisotropic contribution $\textbf{J}_\textbf{1}$,  long-range antiferromagnetic isotropic intra-chain exchange $J_4$ and antiferromagnetic isotropic exchange $J_\bot$ between neighboring ladders without anisotropic contributions. The anisotropic contribution to $J_4$ can be expected to be very small compared to $\textbf{J}_\textbf{1}$ because of the longer Cu--O--O--Cu super-super exchange path. A similar argument holds for $J_\bot$, which is not indicated in Fig.~\ref{LiCu2O2_schematic_view}, because its direction is oriented along the crystallographic \emph{a} axis and so $J_\bot$ is perpendicular to the plane of Fig.~\ref{LiCu2O2_schematic_view}. The first term of the last line of Eq.~\ref{math/4} introduces a possible antisymmetric anisotropic exchange interaction, i.e. a Dzyaloshinskii-Moriya (DM) interaction, within the chains, which in this way has not been considered so far, but will appear to be essential to explain the experimentally observed anisotropy of the linewidth. The last term in Eq.~\ref{math/4} denotes the Zeeman interaction of all spins with the magnetic field.

To evaluate the anisotropic exchange contributions in  ${\cal H}_{\rm int}$, one has to consider the respective bond geometries. For each anisotropic exchange contribution a local coordinate system has to be defined such that the corresponding tensor of anisotropic interaction is diagonal and the sum of diagonal elements equals zero. One of the local axes is directed along the exchange bond. The directions of the two other axes are defined by the symmetry of the local environment. As indicated in Fig.~\ref{LiCu2O2_struct}, for the intra-chain anisotropic exchange interaction $\textbf{J}_{\textbf{2}}$ the local axes are defined as: $x'$ - along the O-O direction within the chain, $y'$ - along the Cu-Cu direction within the chain, and $z'$ - perpendicular to the plane spanned by the Cu-O$_2$ ribbons within the chain. The local axes of the intra-ladder anisotropic exchange between neighboring chains $\textbf{J}_{\textbf{1}}$ are chosen as: $x''$ - along the Cu-Cu direction between neighboring chains, $y''$ - perpendicular to the plane spanned by the Cu-O-Cu bridge between neighboring chains, and $z''$ - perpendicular to $x''$ and $y''$ axes. The unit vectors of the local coordinates in the crystallographic system are given in the Appendix.

For details of second-moment calculations for anisotropic exchange interactions we refer to Ref.~\onlinecite{Rushana}. The intra-chain anisotropic contribution $\textbf{J}_\textbf{2}$ can be adopted from the identical ionic configuration in the related compound LiCuVO$_4$, where we considered the same so called ring-exchange geometry of the Cu-O$_2$ ribbons yielding $J_2^{cc}/k_{\rm B}=-2$\,K.\cite{Krug von Nidda} The remaining intra-ladder anisotropic contribution $\textbf{J}_\textbf{1}$ needs a deeper analysis which will be discussed in the following.

\begin{figure}
\includegraphics[width=0.6\columnwidth]{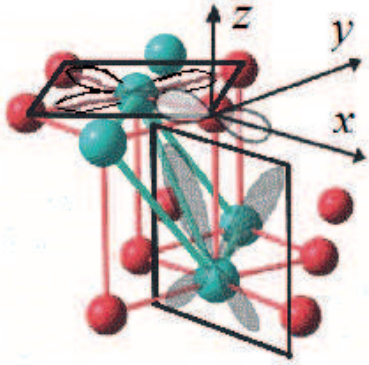}
\caption{(Color online) Schematic pathway of the origin of the anisotropic spin-spin coupling $A_{yy}$
between copper (cyan large spheres) $d_{x^2-y^2}$ (open) states with an excited copper $d_{xz}$ (grey, transparent) state via oxygen (red small spheres) $p_x^c$ (open) states. }\label{LiCu2O2_orbitals_ferro}
\end{figure}

The schematic pathway of the relevant anisotropic spin-spin coupling $\textbf{J}_\textbf{1}$ between two neighboring chains within the same ladder is illustrated in Fig.~\ref{LiCu2O2_orbitals_ferro}.
Here we use local coordinates $x,y,z$ adapted to the conventional rotation of the unperturbed $d$-orbitals neglecting distortion of lattice and any mixing of the wave functions.
We consider the case where the hole ground state $d_{x^2-y^2}$ is coupled with the excited $d_{xz}$ state by spin-orbit interaction (Fig.~\ref{LiCu2O2_orbitals_ferro}). Following this scheme, we estimate the intra-ladder anisotropic exchange parameter $A_{yy}$  according to Ref.~\onlinecite{Bleaney}:
\begin{equation}
\label{math/21}
A_{yy}=\frac{1}{2}\frac{\lambda^2}{\Delta_{x^2-y^2,xz}^2}(\langle x^2-y^2|l_y|xz\rangle)^2 J_{x^2-y^2,xz}
\end{equation}
$J_{x^2-y^2,xz}$ denotes the corresponding isotropic exchange integral, which is significantly larger than $J_1$. A similar expression was obtained earlier in Refs.~\onlinecite{Eremina1,Zakharov}.
To estimate the value of $J_{x^2-y^2,xz}$ we used the formula:
\begin{equation}
\label{math/11}
J_{x^2-y^2,xz}\approx4\frac{t^2_\sigma t^2_\pi}{\Delta_{12}\Delta_{\pi}\Delta_{\sigma}}
\end{equation}
Here we insert $\lambda/k_{\rm B} \approx 913$\,K for the spin-orbit coupling and $\Delta_{x^2-y^2,xz}/k_{\rm B} = (\varepsilon_{5,6}-\varepsilon_{1,2})/k_{\rm B} \approx 13600$\,K for the crystal-field splitting between the respective $d$ states of Cu$^{2+}$ as derived in the Appendix of this paper.

For $\sigma$ and $\pi$ bonds between copper and oxygen, $t_{\sigma}$ and $t_{\pi}$ denote the transfer integrals, $\Delta_{\sigma}$ and $\Delta_{\pi}$ denote the charge-transfer energies. $\Delta_{12}$ corresponds to the charge-transfer energy from one Cu site to the other excited Cu site, which amounts to $\Delta_{12} \approx 7$\,eV.\cite{Papagno}  The ratio of the oxygen-copper transfer parameters $t_{\sigma}$ and $t_{\pi}$ to the charge-transfer energy $\Delta_{\sigma} \approx \Delta_{\pi}$ is known for oxides from studies of the transferred hyperfine interactions as $t^2_{\pi}/\Delta^2_{\pi} \approx t^2_{\sigma}/\Delta^2_{\sigma} \approx 0.077$.~\cite{Walstedt} The oxygen-copper transfer integrals are approximately equal ($t_{\sigma}\approx t_{\pi}$), and according to different estimations their value is about $1.3\leq t_{\sigma}\leq 2.5$\,eV. \cite{Eskes,Hybertsen} Thus, we obtain $74 \leq J_{x^2-y^2,xz} \leq 275$\,meV, which is significantly larger than $J_1 \simeq 3.2$\,meV.

Our estimation of the isotropic exchange interaction parameter between the ground and excited states is quite rough and probably strongly overestimated because of the idealized geometry. Therefore,
to obtain a more realistic value of the anisotropic exchange interaction, we refer to experimental values found for such an exchange geometry in other compounds. In literature the values range from $J=15$\,meV (=174\,K) in Sr$_2$VO$_4$,\cite{Eremin2011}, where the $d_{xy}$ and $p_x$ orbitals exhibit $\pi$-overlapping, to $J=112$\,meV (=1298\,K) in La$_2$CuO$_4$,\cite{Coldea2001} where the overlapping orbitals form $\sigma$ bonds. Using the minimum value $J_{xz,x^2-y^2}=15$~meV in Eq.~\ref{math/21}, we get $A_{yy}/k_{\rm B} \approx 3.5$\,K, which is still significant and cannot be neglected compared to the isotropic exchange.

\begin{figure}
\includegraphics[width=0.8\columnwidth]{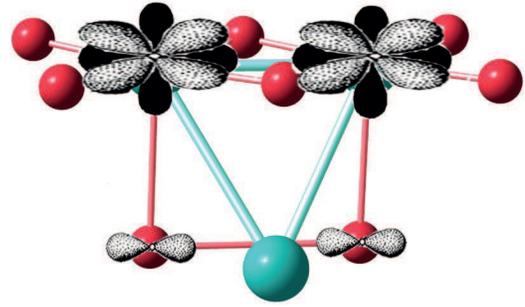}
\caption{(Color online) Possible pathway for realization of antisymmetric anisotropic (DM) exchange coupling $\mathbf{D_{2}}$ between neighboring copper ions within the chains. Cu$^{2+}$ ions (large cyan spheres) with $d_{x^2-y^2}$ (open, dotted) and $d_{xy}$ (full, black) orbitals, O$^{2-}$ (small red spheres) with $p_y$ (open, dotted) orbitals shown on the apical sites} \label{LiCu2O2_orbitals_DM}
\end{figure}

Finally, we indicate a possible exchange path allowing the existence of the DM interaction between neighboring Cu$^{2+}$ ions within the chains. Recently the DM interaction was suggested to be important for stabilization of the spin spiral order,\cite{Furukawa2010,Chen2014} but there is no consensus about its microscopic origin, yet. Furukawa et al.\cite{Furukawa2010} make the inter-layer exchange responsible for a nonzero DM interaction, while Chen amd Hu \cite{Chen2014} suppose that the DM interaction arises within the ladders resulting in a DM vector oriented preferably along the $b$ direction, but at least within the $ab$ plane.

The analysis of our ESR results described below demands a DM vector along the $a$ axis. This could be realized as follows: as the Cu$^{2+}$ ions are built in O$^{2-}$ square pyramids, which in $c$ direction are separated by Cu$^{+}$ planes, we find a Cu$^{2+}$ -- O$^{2-}$ -- O$^{2-}$ -- Cu$^{2+}$ exchange path via the apical oxygen ions giving rise to a DM vector pointing along the $a$ axis, if we neglect the distortions. This exchange path does not have any symmetric counterpart, which would compensate the DM vector. The neighboring chain within the same ladder exhibits the analogue geometry rotated by $180^{\circ}$ giving rise to a DM vector in opposite direction. But as these DM vectors belong to different pairs of Cu$^{2+}$ ions, they do not compensate each other. Due to the admixture of excited orbitals the DM interaction exists, but an estimation of its magnitude is very difficult and demands a deeper theoretical analysis. Hence, we confine ourselves to the experimental determination of the DM contribution in LiCu$_2$O$_2$.

\section{Experimental Results and Discussion}

\begin{figure}
\includegraphics[width=0.8\columnwidth]{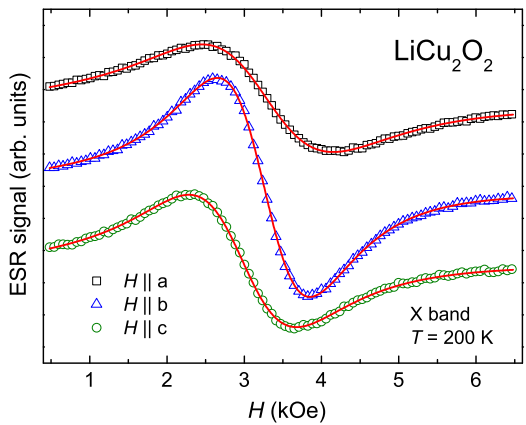}
\caption{(Color online) Typical ESR spectra of LiCu$_2$O$_2$ for the magnetic field applied along the three principal crystallographic axes at $T=200$\,K at $X$-band frequency. Solid lines indicate fits by the field derivative of a Lorentz curve.} \label{LiCu2O2_spectra}
\end{figure}

The untwinned single crystals under investgation were taken from the series of samples grown by solution in the melt described in Ref.~\onlinecite{Svistov2009}. For the ESR measurements they were fixed in Suprasil quartz tubes with paraffin to provide a well defined rotation axis for angular dependent investigations. The ESR measurements were performed in a Bruker ELEXSYS E500-CW spectrometer equipped with continuous-flow He cryostats (Oxford Instruments) at {\it X}- (9.47\,GHz) and {\it Q}-band (34\,GHz) frequencies in the temperature range $4.2 \leq T \leq 300$\,K. Like in earlier reports \cite{Vorotynov,Zvyagin2002} and as shown in Fig.~\ref{LiCu2O2_spectra}, the observed ESR absorption is well described by a single Lorentzian line with resonance field $H_{\rm res}$ and half-width at half maximum linewidth $\Delta H$ within the whole paramagnetic range above $T>35$\,K. Note that the lines with the large linewidth $\Delta H\approx H_{res}$ were fitted including the counter resonance at $-H_{\rm res}$ as described in Ref.~\onlinecite{Joshi2004}.

\begin{figure}
\includegraphics[width=0.9\columnwidth]{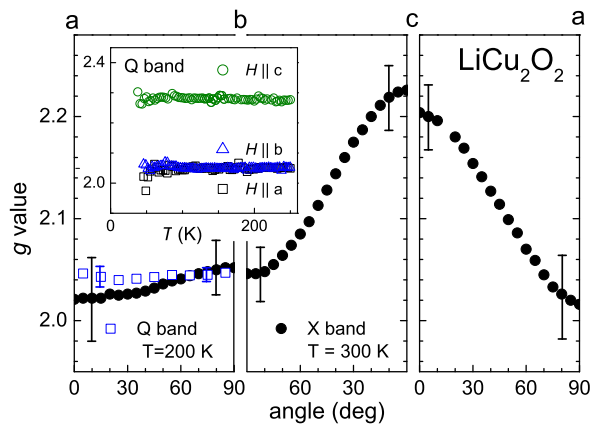}
\caption{(Color online) Angular dependence of the $g$ value in LiCu$_2$O$_2$ for three perpendicular crystallographic planes at $T=300$\,K at $X$-band frequency (solid circles) and for the $ab$ plane at 200~K at $Q$-band frequency (open squares). Inset: Temperature dependence of the $g$ values along the principal axes at $Q$-band frequency.} \label{LiCu2O2_grot_all}
\end{figure}

Fig.~\ref{LiCu2O2_grot_all} shows the angular dependence of the $g$ value at room temperature at $X$-band frequency and partially at $Q$-band frequency for the three principal crystallographic planes.
The $g$ values are independent from temperature for $T \geq 35$\,K, with $g_c = 2.28(1)$ and $g_a=g_b=2.05(1)$ as shown in the inset of Fig.~\ref{LiCu2O2_grot_all}. The observed anisotropy of the $g$ tensor is in agreement with the crystal-field analysis described in the Appendix. Note that due to the point-charge model the calculated $g$ values slightly overestimate the experimental ones.

\begin{figure}
\includegraphics[width=0.9\columnwidth]{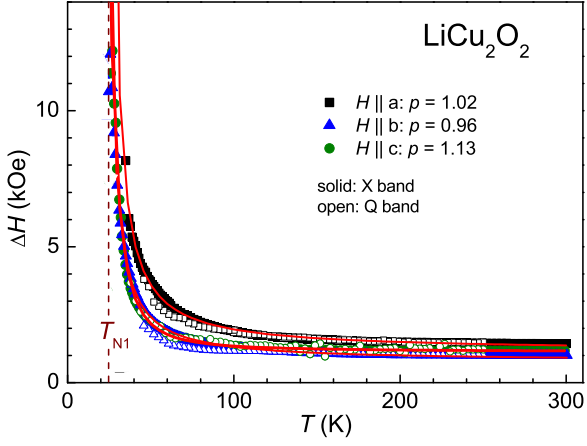}
\caption{(Color online) Temperature dependence of the ESR linewidth in LiCu$_2$O$_2$ with the external magnetic
field applied along the crystallographic axes for $X$- and $Q$-band frequencies. The solid lines indicate fits with a critical divergence $\Delta H = \Delta H_{\infty}+A/(T-T_{\rm N})^p$. The Neel temperature was kept fixed at 24.5~K.}  \label{LiCu2O2_DHvsT_all}
\end{figure}

On decreasing temperature the ESR line strongly broadens and disappears close to the ordering temperature.
This increase of the linewidth towards low temperatures is depicted in Fig.~\ref{LiCu2O2_DHvsT_all} for the field applied along all three principal axes. Fitting the data in terms of a critical law yields different exponents for the three orientations. This results from the competition of different relaxation processes with different temperature dependence and anisotropy. As shown by Oshikawa and Affleck,\cite{Oshikawa2002} the symmetric anisotropic exchange interaction gives rise to a monotonously increasing linewidth with increasing temperature which finally saturates at high temperature. In contrast the DM interaction provokes a divergence of the linewidth on decreasing temperature. In addition, critical behavior may arise because of critical fluctuations close to the N\'{e}el temperature. Due to different anisotropies of these relaxation processes, we cannot scale the temperature dependences of the three main directions on each other.

\begin{figure}
\includegraphics[width=0.9\columnwidth]{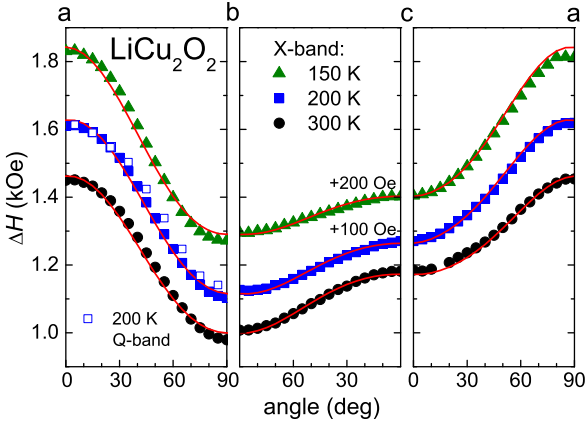}
\caption{(Color online) Angular dependence of the linewidth in LiCu$_2$O$_2$ for three perpendicular crystallographic planes at selected temperatures at $X$-band frequency (solid symbols) and for the $ab$ plane at $T=200$\,K at $Q$-band frequency (open squares). The red solid lines indicate simultaneous fits of all three planes with the parameters given in Table~\ref{Table6}. For clarity, the values of the linewidths at $T = 200$\,K and $T = 300$\,K are shifted by 100\,Oe and 200\,Oe, respectively.}
\label{LiCu2O2_dHrot_all}
\end{figure}

Now we focus on the angular dependence of the linewidth depicted in Fig.~\ref{LiCu2O2_dHrot_all} for $T = 150$, 200 and 300\,K. For all temperatures the maximum of the linewidth is found for $H \parallel a$, the minimum for $H \parallel b$ and an intermediate value for $H \parallel c$ indicating the leading anisotropic exchange contribution to be connected to the $a$ direction. To fit the angular dependencies of the ESR linewidth in LiCu$_2$O$_2$ we used  Eqs.~\ref{math/1}-\ref{math/4}. The isotropic exchange parameters were taken from Ref.~\onlinecite{Masuda_2005}. Hence, as fitting parameters we used the components of the symmetric anisotropic exchange interactions and of the antisymmetric DM interaction $\mathbf{D_2}$ (See Eqs.~\ref{math/4}).

Taking into account the geometry of the exchange bonds, we reduced the number of relevant components to three: in their respective local coordinates the intra-chain interaction $\mathbf{J_2}$ is axial with respect to the $z'$ axis, i.e. $J_2^{z'z'}=-2J_2^{x'x'}=-2J_2^{y'y'}$, the inter-chain interaction $\mathbf{J_1}$ is axial with respect to the $y''$ axis, i.e. $J_1^{y''y''}=-2J_1^{x''x''}=-2J_1^{z''z''}$, and only the DM vector component $D_2^a$ does not vanish.
As one can see, the model provides a good description of the experimental data. The resulting fitting parameters are given in Tab.~\ref{Table6} using the local coordinate systems of the symmetric anisotropic exchange interactions in LiCu$_2$O$_2$ and the crystallographic coordinate system for the DM vector. Note that from the analysis of the angular dependence of the linewidth one obtains the absolute value of the anisotropic exchange parameters. The sign of the anisotropic exchange interaction was derived from the theoretical analysis of the exchange bonds.

\begin{figure}
\includegraphics[width=0.9\columnwidth]{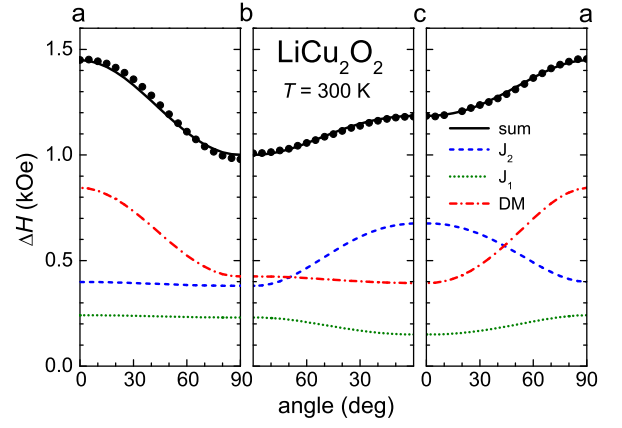}
\caption{(Color online) Angular dependence of the linewidth in LiCu$_2$O$_2$ for three perpendicular crystallographic planes at $T=300$~K at $X$-band frequency (solid symbols) together with the fit contributions from different anisotropic exchange interactions: inter-chain symmetric anisotropic exchange (dotted), intra-chain symmetric anisotropic exchange (dashed), and intra-chain antisymmetric anisotropic exchange $D_2^a$ (dash-dot) and sum (solid).}
\label{LiCu2O2_dHrot_300K}
\end{figure}

Fig.~\ref{LiCu2O2_dHrot_300K} shows the angular dependence of the three contributions separately. While the intra-chain symmetric anisotropic exchange $\mathbf{J_2}$ results in a maximum linewidth for $H \parallel c$ and a nearly constant contribution for $H \perp c$, the inter-chain symmetric anisotropic exchange $\mathbf{J_1}$ leads to a minimum linewidth for $H \parallel c$ and nearly isotropic behavior in the plane $H \perp c$, which can be understood by the superposition of the two inter-chain bonds in the zig-zag ladder. Thus, the symmetric anisotropic exchange contributions $\mathbf{J_1}$ and $\mathbf{J_2}$ together can only result in an effective axial anisotropy of the linewidth with respect to the crystallographic $c$ axis. Therefore, the antisymmetric DM interaction has to be introduced with the only relevant component $D_2^a \neq 0$ which allows describing the observed maximum linewidth for $H \parallel a$.

\begin{table}
\caption{\label{Table6} Parameters of the symmetric anisotropic exchange interactions $\mathbf{J_{1,2}}$ and the antisymmetric Dzyaloshinskii-Moriya (DM) interaction $\mathbf{D_2}$ in the local coordinate system
for the copper ions Cu$^{2+}$ $(S=1/2)$ in LiCu$_2$O$_2$ in units of Kelvin, evaluated at different temperatures. Subscripts correspond to Fig.~\ref{LiCu2O2_schematic_view}}
\begin{tabular}{c c c c c c c c c c c c c c c c c c c}
\hline \hline \\
T (K)   &&& $J_1^{y''y''}$~(K) &&&  $J_2^{z'z'}$~(K) &&& $D_2^a$~(K)
\\ \\ \hline \hline \\
300 &&& 1.20 &&& -1.90 &&& 5.23                   \\
200 &&& 1.20 &&& -1.82 &&& 5.52                  \\
150 &&& 1.28 &&& -1.76 &&& 5.85                  \\
\hline \\
\end{tabular}
\end{table}

The ferromagnetic intra-chain $J_2^{z'z'}$ contribution is of comparable magnitude like in LiCuVO$_4$, where the symmetric anisotropic exchange resulting from the ring geometry in the CuO$_2$ ribbons dominated the spin-spin relaxation. In addition, the intra-ladder contribution is of high importance for the line broadening in LiCu$_2$O$_2$ as predicted by our estimation given above. Interestingly, the DM interaction yields the leading contribution. Here further theoretical efforts will be necessary to understand its origin and possible impact on the still unresolved problem, how to explain the multiferroicity in LiCu$_2$O$_2$.\cite{Sadykov2012}

\section{Summary}

We investigated the spin-spin relaxation of the antiferromagnetic $S=1/2$ quantum spin-ladder compound LiCu$_2$O$_2$ in the paramagnetic regime by means of electron spin resonance. From the anisotropy of the ESR linewidth obtained on untwinned single crystals we were able to extract the symmetric anisotropic exchange contributions resulting from nearest-neighbor super exchange $|J_2^{z'z'}| \sim 1$\,K within the chains and next-nearest neighbor super exchange $|J_1^{y''y''}| \sim 2$\,K within the rungs of the ladders formed by every two neighboring chains. In addition we discovered a sizable intra-chain antisymmetric anisotropic DM contribution $|D_a| \sim 5$\,K which is necessary to describe the observed anisotropy of the linewidth accurately.

Concluding the discussion, let us recall the main microscopic interactions in LiCu$_2$O$_2$ suggested for explanation of the experimental data. The dominant interactions are of isotropic nature: intra-ladder ($J_1$), intra-chain ($J_2$, $J_4$), and inter-ladder exchange interactions ($J_{\perp}$) between the zig-zag ladders located within the $ab$ plane. The inter-plane exchange interactions are at least one order of magnitude smaller.\cite{Masuda_2005} The relativistic interactions in small fields are approximately 5-10 times smaller than the isotropic exchange interactions. The analysis of our ESR data suggests that the strongest of them is the intra-chain antisymmetric anisotropic DM interaction with the DM vector $\mathbf{D_a}$ directed parallel to the crystallographic $a$ direction. The symmetric anisotropic interactions for different exchange paths are found to be 3-5 times smaller. The values of these contributions have the values close to the related chain antiferromagnet LiCuVO$_4$.

The suggested essential intra-chain DM interaction can be important for modeling of the magnetic structure of LiCu$_2$O$_2$ in the magnetically ordered state. In the limit of strong intra-chain DM interaction the chirality vectors of two spiral chains of every zig-zag ladder tend to be antiparallel, because the vectors $\mathbf{D_a}$ for these chains have different signs. Probably, this interaction providing the alternation of chirality vectors explains the absence of spontaneous electrical polarization in the structurally similar magnet NaCu$_2$O$_2$.\cite{Leininger2010}

\section{Acknowledgments}

We thank M.~V.~Eremin for useful discussions concerning the crystal-field analysis.
This work was financially supported by the German Research Foundation (DFG) within the Transregional Collaborative Research Center TRR 80 "From Electronic Correlations to Functionality" (Augsburg, Munich, Stuttgart). L.~E.~S., R.~M.~E., and T.~P.~G. gratefully acknowledge support by the Program of the Steering Committee of the Russian Academy of Sciences.

\appendix

\section{Local Coordinate Systems}

In the unit cell there are two different ladders which both consist of two chains. In the following expressions the upper and lower signs correspond to the directions of the individual vectors for different ladders. In the crystallographic coordinate system $(a,b,c)$  the unit vectors of the local coordinate systems of the intra-chain anisotropic exchange $\textbf{J}_{\textbf{2}}$ read
\begin{equation}
\label{math/J2} \Biggl(
\begin{array}{cccc}
x' & 0.982 & 0 & \mp0.187 \\ y' & 0 & 1 & 0 \\ z'& \pm0.187 & 0 & 0.982
\end{array} \Biggr)
\end{equation}
for first and second ladder, respectively. This means that only the local $y'$ axis coincides with the crystallographic $b$ axis parallel to the Cu$^{2+}$ chains, whereas $x'$ and $z'$ axes are  slightly rotated from  $a$ and $c$ axes, respectively. In the first ladder the unit vectors of the local coordinate system of the intra-ladder anisotropic exchange $\textbf{J}_{\textbf{1}}$ are given by
\begin{equation}
\label{math/J11} \Biggl(
\begin{array}{cccc}
x''& 0.460 & \pm0.463 & 0.758 \\ y''& \mp0.725 & 0.689 & \pm0.018 \\ z''& -0.513 & \mp0.557 & 0.653
\end{array} \Biggr)
\end{equation}
for first and second super-exchange bond, respectively. Analogously, for the second ladder the unit vectors of the local coordinate systems of $\textbf{J}_{\textbf{1}}$ are given by
\begin{equation}
\label{math/J11} \Biggl(
\begin{array}{cccc}
x''& - 0.460 & \pm0.463 & 0.758 \\ y'' & \mp0.725 & -0.689 & \mp0.018 \\ z'' & 0.513 & \mp0.557 & 0.653
\end{array} \Biggr)
\end{equation}

\section{Crystal-Field Analysis}

\begin{figure}
\includegraphics[width=0.6\columnwidth]{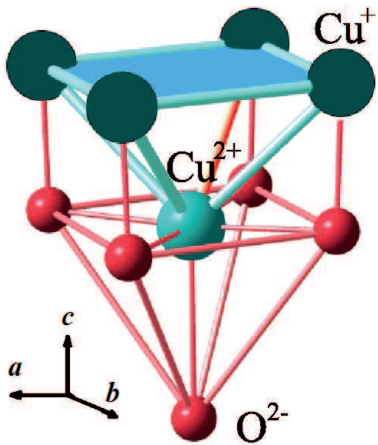}
\caption{(Color online) Local environment of the Cu$^{2+}$ ions in the crystal structure of LiCu$_2$O$_2$.}\label
{LiCu2O2_st}
\end{figure}

\begin{table}
\caption{\label{Table2} Contributions to the crystal-field parameters in
LiCu$_2$O$_2$ at the Cu1$^{2+}$ position (0.124; 1/4; 0.905) in units of Kelvin}
\begin{tabular}{c c c c c c c c c c}
\hline
\hline \\
$B_q^{(k)(K)}$   && point charges and & & point charges     &  sum \\
              && exchange charges  & & ($2.83<r<3.12$\,{\AA}) &        \\
              &&  ($r<2.83$\,{\AA})  & & &     \\ \\ \hline
$B_0^{(2)}$   &&  -30836           & & 3570         &-27266  \\
$B_1^{(2)}$   &&  -1985            & & -450         &-2435  \\
$B_2^{(2)}$   &&  -2629            & & 2760         &131      \\
$B_0^{(4)}$   &&  14407            & & -240         &14167   \\
$B_1^{(4)}$   &&  1487             & & -70          &1417    \\
$B_2^{(4)}$   &&  693              & & -180         &513     \\
$B_3^{(4)}$   &&  1389             & & 25           &1414    \\
$B_4^{(4)}$   &&  -19117           & & -400         &-19517  \\

\hline \hline
\end{tabular}
\end{table}

In LiCu$_2$O$_2$ the Cu$^{2+}$ ions (electronic configuration $3d^9$, spin $S=1/2$) are surrounded by five O$^{2-}$ ions and four Cu$^{+}$ ions. The nearest-neighbor environment of the magnetic Cu$^{2+}$
ion is depicted in Fig.~\ref{LiCu2O2_st}. To calculate the energy-level scheme of Cu$^{2+}$ in LiCu$_2$O$_2$, we start from the following Hamiltonian:
\begin{equation}
\label{math/9}
{\cal H}_0=\lambda(\mathbf{L}\cdot\mathbf{S}) + \sum_{k\,=\,2; \,4}\, \sum_{q\,=\,-k}^{k}B_q^{(k)}\, C_q^{(k)}(\vartheta,\, \varphi)
\end{equation}
The first term corresponds to the spin-orbit coupling. \textbf{S} and \textbf{L} are total spin and orbital moment, respectively. For Cu$^{2+}$ the spin-orbit coupling parameter amounts $\lambda \approx 830$\,cm$^{-1}$.\cite{Abragam} The second term represents the crystal-field operator, where $C_q^{(k)}$ denote the components of the spherical tensor. We use a coordinate system with the Cartesian axes $x$, $y$, and $z$ chosen along the crystallographic axes $a$, $b$, and $c$, respectively. The crystal-field parameters $B_q^{(k)}$ (in eV) were calculated using a superposition model with exchange contributions.\cite{Malkin,Eremin} The relevant overlap integrals were calculated using Hartree-Fock wave functions for Cu$^{2+}$ and O$^{2-}$.\cite{Eremin11} The exchange-charge parameter $G=9.9$ was chosen in accordance with the optical excitation energy $\Delta=1.95$\,eV.\cite{Papagno}

\begin{table}
\caption{\label{Table3} Relative signs of the parameters $B_q^{(k)}$ for Cu2 ($1/2-x$; $1/2+y$; $z-1/2$), Cu3 ($1/2+x$; $y$; $1/2-z$), and Cu4 ($1-x$; $1/2+y$; $1-z$) with respect to the signs for the Cu1 $(x,y,z)$ position in LiCu$_2$O$_2$.}
\begin{tabular}{c c c c cccc c c}
\hline
\hline \\
             & & &   Cu2   &&& & Cu3      && Cu4     \\  \\ \hline
$B_0^{(2)}$  & & &   +     &&& & +        && +               \\
$B_1^{(2)}$  & & &   -     && & & -       && +              \\
$B_2^{(2)}$  & & &   +     &&& & +        && +               \\
$B_0^{(4)}$  & & &   +     && & & +       && +               \\
$B_1^{(4)}$  & & &   -     && & & -          && +               \\
$B_2^{(4)}$  & & &   +     &&& & +            && +               \\
$B_3^{(4)}$  & & &   -     && & & -           && +               \\
$B_4^{(4)}$  & & &   +     && & & +            && +               \\
\hline \hline
\end{tabular}
\end{table}

Using the crystal-field parameters listed in Table~\ref{Table2} for the position Cu$^{2+}$ (0.124; 1/4; 0.906) we obtain the following set of Kramers doublets: $\varepsilon _{1,2}=0$, $\varepsilon _{3,4}=1.077$\,eV, $\varepsilon _{5,6}=1.192$\,eV, $\varepsilon _{7,8} =1.213$\,eV, and $\varepsilon _{9,10}=1.963$\,eV. In the local coordinate system the wave functions read:
\begin{equation}
\label{math/10}
|\varepsilon_n\rangle= \sum_{m_l\,=\,-2}^{+2}\, \sum_{m_S\,=\,\uparrow,\, \downarrow}a_{m_l,\, m_S}^{(n)}\, |m_l,\, m_S\rangle
\end{equation}
The values of the coefficients are given in Tab.~\ref{Table4}. Using these wave functions we calculated the $g$-tensor components $g_{z}=2\langle k_{z}l_{z}+2s_{z}\rangle$, $g_{x}=2\langle k_{x}l_{x}+2s_{x}\rangle$, and $g_{y}=2\langle k_{y}l_{y}+2s_{y} \rangle$, which are equal for all four copper positions: assuming the reduction factors of the orbital momentum due to covalency effects as $k_{x}=k_{y}=k_{z}=0.8$ we obtained $g_z=2.41$, $g_x=2.09$, and $g_y=2.09$.

Note that the energy level scheme derived here, differs from that reported\cite{Huang2011} for the CASSCF/MRCI $d-d$ excitation energies for edge sharing chains of CuO$_4$ plaquettes in LiCu$_{2}$O$_{2}$; $0$ $(d_{xy});$ $1.13/1.43$\,eV $(d_{x^2-y^2});$ $1.58/1.88$\,eV $(d_{xz});$ $1.64/1.94$\,eV $(d_{yz});$ $1.67/1.98$\,eV $(d_{z^2})$, since we have taken into account the contributions to the crystal field from long-distant ligands, which are not negligible.

\begin{table}
\caption{\label{Table4} Coefficients for the Kramers components of ground state $(n=1,2)$ and excited states $(n=3-8)$ in LiCu$_2$O$_2$ at the Cu1 position (0.124; 1/4; 0.906).}
\begin{tabular}{c c c cc cc c| cccccccc}
\hline
\hline \\
$a_{m_l,m_s}^{(1,2)}$&&& $m_s=\downarrow$&&& $m_s=\uparrow$&&&$a_{m_l,m_s}^{(3,4)}$&&&$m_s=\downarrow$&&&$m_s=\uparrow$ \\  \\ \hline
$m_l=2$            &&&  -0.6560        &&& -0.0022       &&&$m_l=2$            &&& -0.3293       &&& -0.3986  \\
$m_l=1$            &&&  0.0449         &&& -0.0453       &&&$m_l=1$            &&& 0.1153        &&& -0.3040 \\
$m_l=0$            &&&  0.0039         &&& 0.0035        &&&$m_l=0$            &&& 0.0403        &&& 0.0359   \\
$m_l=-1$           &&&  -0.0499        &&& -0.0002       &&&$m_l=-1$           &&& 0.6321        &&& 0.0786   \\
$m_l=-2$           &&&  -0.7504        &&& -0.0059       &&&$m_l=-2$           &&& 0.2696        &&& 0.3816   \\
\hline \hline \\
$a_{m_l,m_s}^{(5,6)}$&&& $m_s=\downarrow$&&& $m_s=\uparrow$&&&$a_{m_l,m_s}^{(7,8)}$&&&$m_s=\downarrow$&&&$m_s=\uparrow$ \\  \\ \hline
$m_l=2$            &&&  0.2325         &&& 0.2447         &&&$m_l=2$            &&& -0.4089         &&& 0.1082  \\
$m_l=1$            &&&  -0.926         &&& -0.1967        &&&$m_l=1$            &&& -0.7437         &&& 0.3729 \\
$m_l=0$            &&&  -0.0355        &&& -0.0409        &&&$m_l=0$            &&& 0.0553          &&& -0.1287   \\
$m_l=-1$           &&&  0.5537         &&& -0.6258        &&&$m_l=-1$           &&& 0.1235          &&& -0.0593   \\
$m_l=-2$           &&&  -0.2324        &&& -0.2891        &&&$m_l=-2$           &&& 0.2825          &&& -0.1035   \\
\hline \hline
\end{tabular}
\end{table}

\end{document}